\documentclass{iopart}
\usepackage{bm}        
\usepackage{iopams}
\usepackage{graphicx}
\newcommand{\pa}{\partial}
\begin{document}

\title{On the theory of cavities with point-like perturbations. Part I: General theory}
\author{T.~Tudorovskiy, R.~H\"ohmann, U.~Kuhl, H.-J.~St\"ockmann}
\address{Fachbereich Physik der Philipps-Universit\"at Marburg, D-35032 Marburg, Germany}
\ead{timur.tudorovskiy@physik.uni-marburg.de}

\date{\today}

\begin{abstract}
The theoretical interpretation of measurements of ``wavefunctions'' and spectra in electromagnetic cavities excited by antennas is considered. Assuming that the characteristic wavelength of the field inside the cavity is much larger than the radius of the antenna, we describe antennas as ``point-like perturbations''. This approach strongly simplifies the problem reducing the whole information on the antenna to four effective constants. In the framework of this approach we overcame the divergency of series of the phenomenological scattering theory and justify assumptions lying at the heart of ``wavefunction measurements''. This selfconsistent approach allowed us to go beyond the one-pole approximation, in particular, to treat the experiments with degenerated states. The central idea of the approach is to introduce ``renormalized'' Green function, which contains the information on boundary reflections and has no singularity inside the cavity.
\end{abstract}

\maketitle

\section{Introduction}

Cylindrical microwave cavities with different cross-sections and small height are now widely used as the simplest realization of ``quantum'' (wave) billiards \cite{sri91,ste95,kuh05b}. Indeed for the lowest TM-mode (transversal magnetic) inside of the cavity the transversal component of electric field satisfies the two-dimensional Helmholz equation with Dirichlet boundary condition \cite{jac62}. This problem is mathematically equivalent to the problem for the quantum particle inside a box with hard walls. This boundary problem can not be solved analytically for arbitrary cavity shapes, where the corresponding classical problem is non-integrable and shows chaotic behavior. Microwave cavities are a simple  realization of ``quantum'' chaotic systems. In spite of its experimental simplicity, microwave billiards exhibit general properties of quantum chaotic systems \cite{stoe99}.

Experimentally the cavity is always open, since one has to excite and measure the field inside. Usually one antenna is used for an excitation and the same or a second one for the measurement \cite{stoe99} (see Fig.\ref{FigCavity}). These antennas can be considered as ``probes'', assuming that the influence of antennas on the investigated properties is comparatively small. However this assumption quite often is violated. To understand the data in these cases we have developed a self-consisted theory of the experiment.

The detailed analysis of the antenna construction represents an individual complicated problem \cite{bar03}. Therefore one has to treat antennas phenomenologically to obtain analytical formulas. The standard scattering matrix formalism \cite{mah69,lew91,stoe99} applied to the billiard system does not take properly into account singularities from the sources and contains divergent series. This disadvantage can be overcome if one uses an approach based on ``point perturbations'' of the system \cite{dem88,alb88}. The main idea of the approach is the introduction of the ``renormalized'' Green function instead of the Green function with coincident space arguments.

In Section \ref{sec2} we explain the idea of a point-like perturbation in two-dimensional case and derive the matching condition between wavefunctions inside of the antenna and inside of the cavity. Doing this we basically follow the line of Exner and $\check{\textrm{S}}$eba \cite{exn97}. Ibidem we discuss the physical meaning of the coupling coefficients. In Section \ref{sec3} we generalize the scattering matrix ($S$-matrix) obtained in \cite{exn97} to the case of multiple antennas and estimate coupling coefficients for metallic and dielectric scatterers. Section \ref{sec4} represents a general theory of an experimental wavefunction extraction based on the expression for $S$-matrix obtained in the previous section. The theory is based on the one-pole approximation in the isolated resonances regime. At the end of the Section we show what happens when the ``wavefunction'' is measured in the case of degenerated states.

While for closed cavities one can take into account only a finite number of resonances inside of the experimental range of wavenumbers, for open cavities this approximation fails. In Section \ref{sec5} we show that for an open cavity with broad resonances the transmission coefficient between two antennas is proportional to the Green function of the unperturbed system. This gives the possibility to investigate the spatial structure of the Green function of an open cavity, for example nodal domain distributions \cite{kuh07a,kuh07b}.

In Section \ref{sec6} we consider the spectrum of a cavity with antennas connected by loops. Such a cavity represents a simple model of a quantum graph containing parts of different dimensions.
In Section \ref{sec7} we generalize the model to treat multimode cavities.

\ref{appA} gives the link between the presented approach and the usual $WW^+$ approach where point-like perturbations are treated as $\delta$-functions. In \ref{appB} we write the Green function for the cavity with antennas. Poles of this Green function determine resonances of the system disturbed by the antennas.

\section{\label{sec2}Matching for point-like antenna}

\begin{figure}
\raisebox{1cm}[0pt][0pt]{a)}
\parbox{3.7cm}{\includegraphics[width=3.7cm]{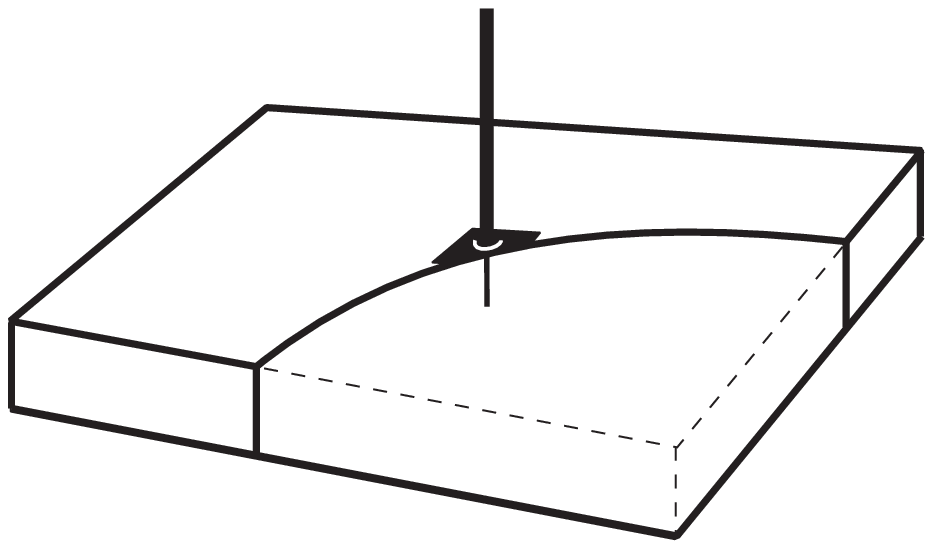}}
\raisebox{1cm}[0pt][0pt]{b)}
\parbox{3.7cm}{\includegraphics[width=3.7cm]{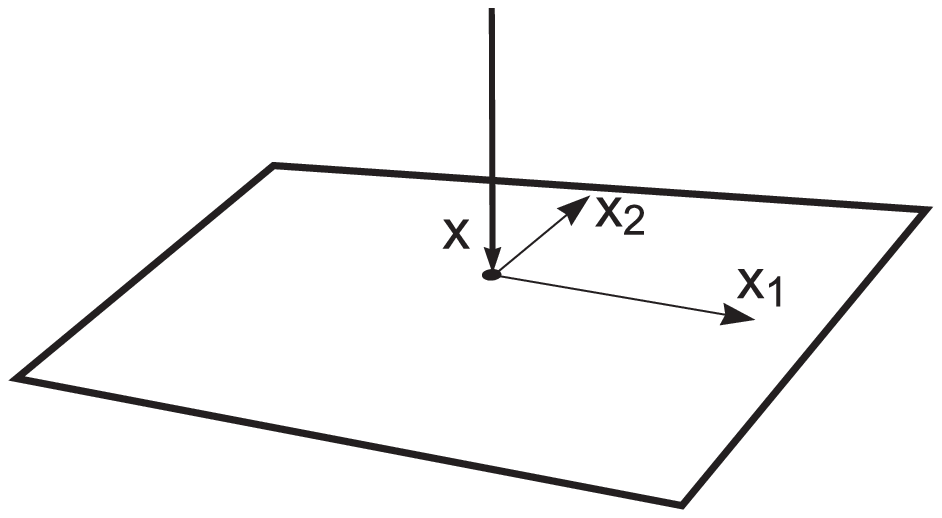}}
\caption{\label{FigCavity} a) The scheme of the physical cavity with antenna. b) Mathematical model of the cavity.}
\end{figure}

We now discuss the general idea of the approach used in Ref.~\cite{exn97} which allows us to treat the presence of scatterers and antennas in a simple way assuming that the wavelength of the field is much larger than the characteristic size of the area of the attachment. Figure \ref{FigCavity} illustrates the principal setup to measure spectra of a cavity. One sees that a quasi-one dimensional waveguide is attached to a cavity at some point. We will assume that the waveguide is straight and one can separate variables inside of it and define a scalar longitudinal wavefunction depending on the variable inside the waveguide. Furthermore we assume that the height of the cavity is small and the distance between transversal wavenumbers is large. Thus it is natural to suppose that the one-dimensional longitudinal wavefunction in the waveguide relates to the longitudinal wavefunction in the cavity in such a way that only some general and energy independent properties of the matching are important. Since the detailed construction of the matching usually is quite complicated,
the only restriction applied is that the final operator describing the coupling between antenna and cavity should be linear and Hermitian.

We substantially follow paper \cite{exn97}. Basically the idea of the approach is to construct a self-adjoint extension of the Hamiltonian in the presence of point-like perturbations. In various situations self-adjoint extensions were considered \cite{alb88,vor06} and related problems were studied \cite{alb96,bru03}.

Let us consider the matching of a quasi one-dimensional antenna to a cavity.
We denote by $\varphi_1(x),\,-\infty<x\leq 0$ the longitudinal wavefunction inside the one-dimensional antenna and by $\varphi_2(\mathbf{r}),\,\mathbf{r}=\{x_1,x_2\}$ a function inside the cavity. Thus $\varphi=\{\varphi_1(x),\varphi_2(\mathbf{r})\}$ is a function on the space consisting of two parts: cavity and antenna. We assume that the antenna is coupled at the end $x=0$ to the point $\mathbf{r}=\mathbf{R}$ of the cavity. We define the operator $H$ which acts on the function $\varphi$ in the following way:
\begin{equation}
H\varphi=\left(\begin{array}{l}-\varphi_1''(x) \\ -\Delta\varphi_2(\mathbf{r})\end{array}\right),
\label{OperDef}
\end{equation}
where $\varphi_1''(x)=d^2\varphi_1(x)/dx^2$ and $\Delta$ is the two-dimensional Laplacian. Since for TEM-mode in the waveguide and the lowest TM-mode in the cavity thresholds do not exist we did not consider them in Eq. \eref{OperDef}. The generalization to higher transversal modes in the cavity will be presented in the Section \ref{sec7}.

In the stationary case we look for the solution of the spectral problem
\begin{equation}
H\varphi=k^2\varphi,
\label{SpecProb}
\end{equation}
where $k$ is the total wavenumber.

Now we turn to the question how to define operator $H$ to be Hermitian. This is equivalent to the introduction of proper boundary matching conditions for the functions $\varphi_1(x)$ and $\varphi_2(\mathbf{r})$.

In order to avoid divergent integrals at the origin first we consider the cavity with a small pricked ring of radius $\epsilon$ with a center at the point $\mathbf{r}=\mathbf{R}$. We call this area $\Omega_\epsilon$. Following the general procedure \cite{lan65c,dem88,exn97} we consider two quadratic forms
\begin{eqnarray}
\left<\varphi^+ H\psi\right>_\epsilon\equiv
-\int_{-\infty}^0 dx\bar\varphi_1 \psi_1''-\int_{\Omega_\epsilon} d^2r\bar\varphi_2 \Delta\psi_2,\nonumber\\
\left<\psi^+ H\varphi\right>_\epsilon\equiv
-\int_{-\infty}^0 dx\bar\psi_1 \varphi_1''-\int_{\Omega_\epsilon} d^2r\bar\psi_2 \Delta\varphi_2,
\label{rhoforms}
\end{eqnarray}
where bar means complex conjugation, superscript `$+$' denotes hermitian conjugation and angle brackets mean integration. Then consider the difference
\begin{eqnarray}
\left<\varphi^+ H\psi\right>_\epsilon-\overline{\left<\psi^+ H\varphi\right>_\epsilon}=
\int_{-\infty}^0 dx(\psi_1\bar\varphi_1'-\bar\varphi_1 \psi_1')'+\nonumber\\
+\int_{\Omega_\epsilon}\nabla(\psi_2\nabla\bar\varphi_2-\bar\varphi_2 \nabla\psi_2) d^2r=
[\psi_1\bar\varphi_1'-\bar\varphi_1 \psi_1']_{x=0}-\nonumber\\
-\epsilon\int_0^{2\pi} d\vartheta [\psi_2 (\pa\bar\varphi_2/\pa \rho)-\bar\varphi_2 (\pa\psi_2/\pa \rho)]_{\rho=\epsilon}.
\label{difference}
\end{eqnarray}
In the last equality we introduced cylindrical coordinates $\rho,\vartheta:\,\{x_1=\rho\cos\vartheta,x_2=\rho\sin\vartheta\}$, $\rho=|\mathbf{r}-\mathbf{R}|$, and assumed 
\begin{eqnarray}
[\psi_1\bar\varphi_1'-\bar\varphi_1 \psi_1']_{x=-\infty}=0, \nonumber\\
\left[\psi_2\nabla_n\bar\varphi_2-\bar\varphi_2 \nabla_n\psi_2\right]_\sigma=0,
\end{eqnarray}
where $\sigma$ is the physical boundary of the cavity and $\nabla_n$ is the derivative along the outer normal to the boundary. This is equivalent to the assumption that the functions or their normal derivatives are zero at the boundary (Dirichlet or Neumann boundary conditions or their linear combination).

Assuming that functions $\varphi_2,\,\psi_2$ have the same kind of singularities for $\mathbf{r}\to\mathbf{R}$ as the Green function for the two-dimensional plane \cite{duf01,gra80}
\begin{eqnarray}
G(\mathbf{r},\mathbf{R};k^2)=\frac{i}{4}H_0^{(1)}(k|\mathbf{r}-\mathbf{R}|),\nonumber\\
G(\mathbf{r}\to\mathbf{R},\mathbf{R};k^2)\to
-\frac{1}{2\pi}\left(\ln\frac{k|\mathbf{r}-\mathbf{R}|}{2}+\gamma\right)+\frac{i}{4},
\label{G2Dplane}
\end{eqnarray}
where $H_0^{(1)}$ is the Hankel function of the first kind and $\gamma\simeq 0.5772$ is Euler constant, we can write asymptotic expansions for $\rho\to 0$:
\begin{eqnarray}
\varphi_2(\rho,\varphi)=(2\pi)^{-1}L_0^{\varphi_2}\ln(\rho)+L_1^{\varphi_2}+f_\varphi(\rho,\varphi), \nonumber\\
\psi_2(\rho,\varphi)=(2\pi)^{-1}L_0^{\psi_2}\ln(\rho)+L_1^{\psi_2}+f_\psi(\rho,\varphi),
\label{assol}
\end{eqnarray}
where $L_0^{\varphi_2},\,L_1^{\varphi_2},\,L_0^{\psi_2},\,L_1^{\psi_2}$ are constants.
We have written the factor $(2\pi)^{-1}$ in order to simplify the further calculations.
Here $\rho$ is calculated in arbitrary length units. Though $\ln(\rho)$ depends on the choice of the length unit, the sum of the first two terms in expansions \eref{assol} is invariant. We will discuss this feature later.

Assuming that $f_{\varphi,\psi}(\rho)\to 0$ and  $|\pa f_{\varphi,\psi}/\pa \rho|<\textrm{const}$ for $\rho\to 0$, we get
\begin{eqnarray}
\lim_{\epsilon\to 0}
\epsilon\int_0^{2\pi} d\vartheta [\psi_2 (\pa\bar\varphi_2/\pa \rho)-\bar\varphi_2 (\pa\psi_2/\pa \rho)]_{\rho=\epsilon}
=\bar L_0^{\varphi_2}L_1^{\psi_2}-\bar L_1^{\varphi_2}L_0^{\psi_2}.
\end{eqnarray}
Finally in the limit $\epsilon\to 0$ equality \eref{difference} takes the form
\begin{eqnarray}
\left<\varphi^+ H\psi\right> - \overline{\left<\psi^+ H\varphi\right>}
\equiv
\lim_{\epsilon\to 0}
\left[\left<\varphi^+ H\psi\right>_\epsilon - \overline{\left<\psi^+ H\varphi\right>_\epsilon}\right]
=\nonumber\\=
\left[\bar\varphi_1'\psi_1-\bar\varphi_1 \psi_1'\right]_{x=0}
+\bar L_1^{\varphi_2}L_0^{\psi_2}-\bar L_0^{\varphi_2}L_1^{\psi_2}.
\label{DiffQForms}
\end{eqnarray}
Let us introduce vectors $\eta_\varphi=(\varphi_1'(0),L_1^{\varphi_2})$, $\zeta_\varphi=(\varphi_1(0),L_0^{\varphi_2})$, $\eta_\psi=(\psi_1'(0),L_1^{\psi_2})$, $\zeta_\psi=(\psi_1(0),L_0^{\psi_2})$ and rewrite the last equality in the form
\begin{equation}
\left<\varphi^+ H\psi\right> - \overline{\left<\psi^+ H\varphi\right>}=
\left<\eta_\varphi,\zeta_\psi\right>-\left<\zeta_\varphi,\eta_\psi\right>,
\label{formsdifference}
\end{equation}
where $\langle\cdot,\cdot\rangle$ denotes the scalar product of two vectors.
Since the operator $H$ is linear we write
\begin{equation}
\eta_\varphi=\Lambda \zeta_\varphi,\quad \eta_\psi=\Lambda \zeta_\psi
\label{CouplConst}
\end{equation}
and get that matrix $\Lambda$ has to be Hermitian. Thus the general matching condition reads
\begin{equation}
\left(\begin{array}{l}
\varphi_1'(0) \\ L_1^{\varphi_2}
\end{array}\right)=
\left(\begin{array}{ll}
A & B \\
C & D
\end{array}\right)
\left(\begin{array}{l}
\varphi_1(0) \\ L_0^{\varphi_2}
\end{array}\right)
\label{ant-matching},
\end{equation}
where $A,\,D$ are real and $C=\bar{B}$. As it is assumed that $H$ is $k$-independent, the constants $A,\,B,\,C$ and $D$ are $k$-independent as well. From the last formula one can see that for $B=C=0$ antenna and cavity are decoupled, which means there is no mixing between functions $\varphi_1,\,\varphi_2$: $\varphi_1'(0)=A\varphi_1(0)$, $ L_1^{\varphi_2}=D L_0^{\varphi_2}$. Thus we conclude that $B,\,C$ are coupling constants, $A$ describes the boundary condition at the end of the antenna and $D$ describes the scattering strength of the antenna.
$A=\infty$ corresponds to $\varphi_1(0)=0$ (Dirichlet boundary condition) and $A=0$ corresponds to $\varphi_1'(0)=0$ (Neumann boundary condition).

Let us now prove the invariance on the length scale of the expression \eref{assol}. From the matching condition \eref{ant-matching} we have
\begin{equation}
L_1^{\varphi_2}=D L_0^{\varphi_2} + C \varphi_1(0).
\end{equation}
Substituting this relation in \eref{assol} we get
\begin{eqnarray}
\varphi_2(\rho,\varphi)=L_0^{\varphi_2}[(2\pi)^{-1}\ln(\rho)+D]+C \varphi_1(0)+f_\varphi(\rho,\varphi).
\label{assol1}
\end{eqnarray}
Let us introduce the scattering length $\beta$ by the following equality
\begin{equation}
D=-\frac{1}{2\pi}\ln\beta.
\label{scldef}
\end{equation}
Then \eref{assol1} takes the form
\begin{eqnarray}
\varphi_2(\rho,\varphi)=\frac{L_0^{\varphi_2}}{2\pi}\ln\frac{\rho}{\beta}+C \varphi_1(0)+f_\varphi(\rho,\varphi),
\label{assol2}
\end{eqnarray}
which is invariant. The definition of the scattering length \eref{scldef} corresponds to the condition that the wavefunction is zero at the distance $\beta$ from the perturbation point when $B=C=0$. It differs from the definition given in \cite{dem88}.

\section{\label{sec3}$S$-matrix for a cavity with multiple antennas}

Here we consider the case with $\nu$ antennas attached to a cavity at points $\mathbf{R}_1,\,\mathbf{R}_2,\ldots,\mathbf{R}_\nu$.
We look for the solution in the form
\begin{equation}
\label{ManyAntennasStat}
\begin{array}{l}
\varphi_{1i}(x)=a_i e^{ikx}+b_i e^{-ikx},\\
\varphi_2(\mathbf{r})=\sum_{i=1}^\nu F_i G(\mathbf{r},\mathbf{R}_i;k^2),
\end{array}
\end{equation}
where $\varphi_{1i}(x)$ is the solution inside the $i$-th antenna, $\varphi_2(\mathbf{r})$ is the field inside the cavity, $G(\mathbf{r},\mathbf{R}_i;k)$ is the Green function in the cavity
and $F_i$ are unknown coefficients which have to be defined.

The Green function $G(\mathbf{r},\mathbf{R}_i;k^2)$ has singularities at the points $\mathbf{r}=\mathbf{R}_i$. Thus in the neighborhood of $\mathbf{R}_i$ we write
\begin{equation}
G(\mathbf{r}\to\mathbf{R}_i,\mathbf{R}_i;k^2)\to-\frac{1}{2\pi}\ln|\mathbf{r}-\mathbf{R}_i|+\xi(\mathbf{R}_i;k^2),
\end{equation}
where $\xi(\mathbf{R}_i;k^2)$ represents the \textit{renormalized} Green function. It plays the key role in the following calculations. Numerically the \textit{renormalized} Green function can be calculated by means of the boundary element method \cite{shu97}.

From the definition we see that $\xi(\mathbf{R}_i;k^2)$ depends on the chosen system of units. However all equations determining physical observable quantities will of course be invariant.

Then we define generalized boundary values $L_{0i}^{\varphi_2},\,L_{1i}^{\varphi_2}$ at the matching points
\begin{eqnarray}
\varphi_{2i}(\mathbf{r}\to\mathbf{R}_i)\to(2\pi)^{-1}L_{0i}^{\varphi_2}\ln(|\mathbf{r}-\mathbf{R}_i|)
+L_{1i}^{\varphi_2}, \nonumber\\
L_{0i}^{\varphi_2}=-F_i, \nonumber\\
L_{1i}^{\varphi_2}=F_i\xi(\mathbf{R}_i;k)+
\sum_{j\neq i}F_j G(\mathbf{R}_i,\mathbf{R}_j;k^2).
\label{L0L1}
\end{eqnarray}
Using matching condition \eref{ant-matching}, we get
\begin{equation}
\fl
\begin{array}{r}
\left(\begin{array}{l} ik(a_i-b_i) \\ F_i\xi(\mathbf{R}_i;k^2)+
\sum_{j\neq i}F_jG(\mathbf{R}_i,\mathbf{R}_j;k^2)
\end{array}\right)
=\left(\begin{array}{ll} A_i & B_i \\ C_i & D_i \end{array}\right)
\left(\begin{array}{l}
a_i + b_i \\ -F_i
\end{array}\right).
\end{array}
\label{AntsCavMatching}
\end{equation}
These equations can be rewritten in a matrix form
\begin{eqnarray}
BF=(A-ik)a+(A+ik)b, \label{eq01}\\
(\hat{G}+D) F=C(a+b) \label{eq02},
\end{eqnarray}
where vectors $F,\,a,\,b$ have elements $F_i,\,a_i,\,b_i$ respectively, matrix $\hat G$ has elements
\begin{eqnarray}
\hat G_{ij}=\hat G_{ij}(\mathbf{R}_1,\ldots,\mathbf{R}_\nu;k^2)=\left\{\begin{array}{ll}
\xi(\mathbf{R}_i;k^2), & i=j, \\
G(\mathbf{R}_i,\mathbf{R}_j;k^2), & i\neq j,
\end{array}\right.
\end{eqnarray}
and diagonal matrices $A,\,B,\,C,\,D$ have elements $A_i,\,B_i,\,C_i,\,D_i$ on diagonals respectively. Substituting $F$ from \eref{eq02} into \eref{eq01} we get
\begin{equation}
(M-ik)a+(M+ik)b=0,
\end{equation}
where $M=A - B(\hat G+D)^{-1}C$. Thus
\begin{equation}
b=Sa,
\end{equation}
where the $S$-matrix has the form
\begin{eqnarray}
S=-\frac{M-ik}{M+ik}.
\label{Smatrix}
\end{eqnarray}
The Green function of a closed cavity is real, thus from Eq. \eref{Smatrix} one sees that the $S$-matrix is unitary.
Using the equality
\begin{eqnarray}
\frac{M-ik}{M+ik}-\frac{A-ik}{A+ik}=
2ik\frac{1}{M+ik}(M-A)\frac{1}{A+ik},
\end{eqnarray}
we represent $S$ in the following way:
\begin{eqnarray}
S=
-\frac{A-ik}{A+ik}+
\frac{2ikB}{A+ik}\left[\hat G+D-\frac{BC}{A+ik}\right]^{-1}\frac{C}{A+ik}.
\label{Smatrix1}
\end{eqnarray}
For the case of a single antenna formula \eref{Smatrix1} was obtained in \cite{exn97}.

From \eref{Smatrix1} we see that poles of $S$-matrix $k=k(\mathbf{R}_1,\ldots,\mathbf{R}_\nu)$ are the complex solutions of the equation
\begin{equation}
\det\left[\hat G(\mathbf{R}_1,\ldots,\mathbf{R}_\nu;k^2)+D-\frac{BC}{A+ik}\right]=0.
\label{Spoles}
\end{equation}
In the limit of a closed system
($B,\,C\to 0$) we get the equation determining the spectrum of the system with only scatterers inside
\begin{equation}
\det[\hat G(\mathbf{R}_1,\ldots,\mathbf{R}_\nu;k^2)+D]=0.
\label{SpecScat}
\end{equation}
Eq. \eref{SpecScat} shows that in the limit $D\to\infty$, poles of the system approach the poles of the unperturbed system.

\subsection{One-dimensional cavities}
If an antenna is attached to a narrow waveguide with a width comparable to the wavelength, then the cavity can be considered as the one-dimensional one. The $S$-matrix for such a cavity can be calculated analogously to two-dimensional case. We introduce the operator inside the cavity
\begin{equation}
H\varphi=\left(\begin{array}{l}
-\varphi_1''(x) \\
(-d^2/dx_1^2 + k_2^2)\varphi_2(x_1)
\end{array}\right),
\label{OperDef1d}
\end{equation}
where $k_2^2$ is a threshold from the quantization in $x_2$ direction. We are looking for the solution of the spectral problem \eref{SpecProb}.
We denote $X$ the position of the antenna attachment, $X_l,\,X_r$ are the coordinates of the ends of the cavity, $X_l<X_r$. Calculating the difference of two quadratic forms we get
\begin{eqnarray}
\left<\varphi^+H\psi\right> - \overline{\left<\psi^+H\varphi\right>}\equiv\int_{-\infty}^0 dx(-\bar\varphi_1 \psi_1''+\psi_1\bar\varphi_1'')+\nonumber\\
+\int_{X_l}^{X-\epsilon}dx_1(-\bar\varphi_2 \psi_2''+\psi_2\bar\varphi_2'')
+\int_{X+\epsilon}^{X_r}dx_1(-\bar\varphi_2 \psi_2''+\psi_2\bar\varphi_2'')=\nonumber\\
=[\psi_1\bar\varphi_1'-\bar\varphi_1 \psi_1']_{x=0}+[\psi_2\bar\varphi_2'-\bar\varphi_2\psi_2']\Bigr|_{X+\epsilon}^{X-\epsilon}
=\left<\eta_\varphi,\zeta_\psi\right>-\left<\zeta_\varphi,\eta_\psi\right>,
\label{formsdif1d}
\end{eqnarray}
where $\epsilon\to 0$,
\begin{eqnarray}
\zeta_\varphi=(\varphi_1(0),\varphi_2(X-\epsilon),\varphi_2'(X+\epsilon)),\nonumber\\
\eta_\varphi=(\varphi_1'(0),\varphi_2'(X-\epsilon),\varphi_2(X+\epsilon)),\nonumber\\
\zeta_\psi=(\psi_1(0),\psi_2(X-\epsilon),\psi_2'(X+\epsilon)),\nonumber\\
\eta_\psi=(\psi_1'(0),\psi_2'(X-\epsilon),\psi_2(X+\epsilon)).
\end{eqnarray}
Then we write $\eta_{\varphi,\psi}=\Lambda\zeta_{\varphi,\psi}$. From \eref{formsdif1d} $\Lambda$ is Hermitian. Usually one assumes that $\varphi(X-\epsilon)=\varphi(X+\epsilon)=\varphi(X)$, thus
\begin{equation}
\Lambda=\left(\begin{array}{lll}
A & B & 0 \\
C & D & 1 \\
0 & 1 & 0
\end{array}\right).
\end{equation}
This gives
\begin{eqnarray}
\varphi_1'(0)=A\varphi_1(0)+B\varphi_2(X),\\
\varphi_2'(X-\epsilon)-\varphi_2'(X+\epsilon)=C\varphi_1(0)+D\varphi_2(X).
\label{derj}
\end{eqnarray}
For a cavity with a scatterer inside ($B=C=0$) we get
\begin{equation}
\varphi_2'(X-\epsilon)-\varphi_2'(X+\epsilon)=D\varphi_2(X).
\label{delta-matching}
\end{equation}
Exactly the same condition holds for the derivatives of the solution of the following equation with  true delta-scatterers
\begin{equation}
[-d^2/dx^2-D\delta(x-X)]\varphi_2=(k^2-k_2^2)\varphi_2.
\end{equation}
We have mapped the problem of a 1D cavity with a point-like scatterer to a Schr\"odinger equation with a $\delta$-potential. Such a mapping is possible in one-dimensional system only.

From the asymptotic
\begin{equation}
\varphi_2(x_2\to X)\to L_0^{\varphi_2}|x_2-X|+L_1^{\varphi_2},
\end{equation}
the condition \eref{delta-matching} and the assumption that $L_0^{\varphi_2}\beta+L_1^{\varphi_2}=0$ we find $D=2/\beta$, where $\beta$ is a scattering length.

Let us now construct the $S$-matrix for a number of antennas attached to a cavity at the points $X_1,\,X_2,\ldots,X_\nu$. Analogously to \eref{ManyAntennasStat}, \eref{AntsCavMatching} we write

\begin{equation}
\label{ManyAntennasStat1D}
\begin{array}{l}
\varphi_{1i}(x)=a_i e^{ikx}+b_i e^{-ikx},\\
\varphi_2(x_2)=\sum_{i=1}^\nu F_i G(x_2,X_i;k_1^2),
\end{array}
\end{equation}
where $k_1^2=k^2-k_2^2$ and
\begin{equation}
\label{AntsCavMatching1D}
\begin{array}{l}
\varphi_{1i}'(0)=A_i\varphi_{1i}(0)+B_i\varphi_{2i}(X_i),\\
\varphi_{2}'(X_i-\epsilon)-\varphi_{2}'(X_i+\epsilon)=C_i\varphi_{1i}(0)+D_i\varphi_{2}(X_i),
\end{array}
\end{equation}
where $G(x_2,X;k_1^2)$ is the Green function of the closed one dimensional cavity.
Using the identity
\begin{equation}
\lim_{\epsilon\to 0}\left[\frac{d}{dx_2}G(x_2,X;k_1^2)\right]_{X+\epsilon}^{X-\epsilon}=1,
\end{equation}
we find analogously to \eref{Smatrix1}
\begin{eqnarray}
S=
-\frac{A-ik}{A+ik}+
\frac{2ikB}{A+ik}\left[G^{-1}-D-\frac{BC}{A+ik}\right]^{-1}\frac{C}{A+ik},
\label{Smatrix1D}
\end{eqnarray}
where
\begin{equation}
G_{ij}=G(X_i,X_j;k_1^2).
\end{equation}

On the contrary to \eref{SpecScat} poles of the closed system ($B,C\to 0$) tend to the unperturbed ones when $D\to 0$.

\subsection{Example: Single antenna attached to an infinite cavity}
One can imagine a Gedankexperiment where an antenna is attached to an infinitely large cavity. In such a cavity there are no reflections from the walls. Experimentally this situation can be realized if one places a perfect absorber along the boundary of the cavity. In terms of the cross-section such a resonator can be considered as a two-dimensional plane.

Let us consider a single antenna attached to a plane at the point $\mathbf{R}$. In this case the unique element of $S$-matrix $S_{11}$ is equal to the reflection coefficient and $A,\,B,\,C,\,D$ are scalar constants.
From \eref{Smatrix1} we find
\begin{equation}
S_{11}=-\frac{A-ik}{A+ik}+\frac{2ik B C}{(A+ik)^2}\left[\xi(\mathbf{R};k^2)+D-\frac{BC}{A+ik}\right]^{-1},
\label{s11}
\end{equation}
where
\begin{equation}
\xi(\mathbf{R};k^2)=-\frac{1}{2\pi}\left(\ln\frac{k}{2}+\gamma\right)+\frac{i}{4}.
\end{equation}
The last equality follows from \eref{G2Dplane}. In the limit of a closed cavity the pole of $S_{11}$ is
\begin{equation}
k=2ie^{-\gamma}/\beta\simeq 1.123 i/\beta,
\label{PoleScatPlane}
\end{equation}
where $\beta$ is the scattering length. The eigenvalue of the problem \eref{SpecProb} is $k^2=-4e^{-2\gamma}\beta^{-2}\simeq-1.261\beta^{-2}$.
This eigenvalue corresponds to the eigenstate induced by a point-like scatterer with a wavenumber below the threshold. This means that a point-like scatterer in two-dimensional case can be always treated ``quantum-mechanically'' as an attractive well. Since in electromagnetic cavities the threshold for the lowest transversal mode is equal to zero, this pole can be observed only for higher transversal modes.

\subsection{\label{scatlest}Estimation of the scattering length}
In the previous example we showed that a point-like perturbation can be always treated as an attractive well. Nevertheless using it one can model not only dielectric but also metallic scatterers. The apparent contradiction is explained by the fact that in the long-wave limit the whole information about the scatterer is reduced to its scattering length which can be the same for the attractive as well as for the repulsive scatterer. This description fails whenever two scatterers are close to each other or close to the boundary (compare \cite{bru05}). Below we estimate a scattering length for attractive and repulsive scatterers.

Let us consider a cylindrical cavity with radius $R$ and a cylindrically symmetric scatterer of radius $r_0$ whose center coincides with the centrum of the cavity. We shall treat the scatterer in a quantum-mechanical way as a potential and shall consider states with zero angular momentum.

We put the origin of the frame in the center of the cavity.
We denote the wavefunctions inside and outside of the scatterer by $\psi_{in}(r;k)$ and $\psi_{out}(r;k)$, respectively. Here $r=|\mathbf{r}|$. Then
\begin{equation}
\psi_{out}(r;k)=c_1(k) J_0(kr)+c_2(k) N_0(kr),
\end{equation}
where $J_0$ and $N_0$ are Bessel and Neumann functions of zero order. At the boundary of the cavity $\psi_{out}(R;k)=0$. This gives $c(k)=c_2(k)/c_1(k)=-J_0(kR)/N_0(kR)$.
 Equating logarithmic derivatives at the boundary of the scatterer we find
\begin{equation}
\frac{r_0\psi_{in}'(r_0;k)}{\psi_{in}(r_0;k)}=\frac{-kr_0[J_1(kr_0)+c(k) N_1(kr_0)]}{J_0(kr_0)+c(k) N_0(kr_0)},
\label{ExactMatch}
\end{equation}
where $\psi_{in}'=\pa\psi_{in}/\pa r$. We assume that $kr_0\ll 1$. Then one can put $k=0$ on the left hand-side of \eref{ExactMatch} and use the asymptotic expansions $J_0(x)=1-x^2/4+O(x^4)$, $N_0(x)=(2/\pi)[\ln(x/2)+\gamma]+O(x^2\ln(x))$, $J_1(x)=x/2+O(x^3)$, $N_1(x)=-2/(\pi x)+O(x\ln(x))$ \cite{gra80}.
This gives
\begin{equation}
\frac{r_0\psi_{in}'(r_0;0)}{\psi_{in}(r_0;0)}\simeq\frac{c(k)}{\pi/2+c(k)[\ln(kr_0/2)+\gamma]}.
\end{equation}
From the last equation we get the condition
\begin{equation}
c(k)=\frac{\pi/2}{-\ln(kr_0/2)-\gamma+\psi_{in}(r_0;0)/[r_0\psi_{in}'(r_0;0)]}.
\label{qc1}
\end{equation}

Let us consider what happens if we simulate the scatterer by a point-like perturbation. Using the following equalities
\begin{eqnarray}
G(\mathbf{r},0;k^2)=-\frac{1}{4}\left[J_0(kr)/c(k)+N_0(kr)\right],\nonumber\\
\xi(0;k^2)=-\frac{1}{4c(k)}-\frac{1}{2\pi}\left(\ln\frac{k}{2}+\gamma\right),
\end{eqnarray}
we find from Eqs. \eref{scldef}, \eref{SpecScat}
\begin{eqnarray}
\xi(0;k^2)+D=-\frac{1}{4c(k)}-\frac{1}{2\pi}\left(\ln\frac{k\beta}{2}+\gamma\right)=0.
\label{cirspec}
\end{eqnarray}
This gives the following quantization condition
\begin{eqnarray}
c(k)=-\frac{\pi/2}{\ln(k\beta/2)+\gamma}.
\label{qc2}
\end{eqnarray}
Comparing Eqs. \eref{qc1} and \eref{qc2} we find
\begin{equation}
\beta=r_0\exp\left(-\frac{\psi_{in}(r_0;0)}{r_0\psi_{in}'(r_0;0)}\right).
\label{Destim}
\end{equation}
Let us discuss the last result. First it shows that a singular point-like scatterer can model an attractive as well as a metallic scatterer. For the metallic scatterer of the radius $r_0$ we find $\psi_{in}(r_0)=0$ and $\beta=r_0$ in accordance with \eref{assol2}.

Let us consider now an ``attractive scatterer''. For simplicity we consider a potential well of constant depth $-u_0$. Then $\psi_{in}(r;k)=a J_0(\varkappa r)$, $\varkappa=(u_0+k^2)^{1/2}$, $\psi_{in}(r_0;0)/[r_0\psi_{in}'(r_0;0)]=-J_0(\sqrt{u_0}r_0)/[\sqrt{u_0} r_0 J_1(\sqrt{u_0} r_0)]$. If $\sqrt{u_0} r_0\ll 1$ then
\begin{equation}
\frac{\psi_{in}(r_0;0)}{r_0\psi_{in}'(r_0;0)}\simeq-\frac{2}{u_0 r_0^2}.
\label{festim}
\end{equation}
Substituting \eref{festim} in \eref{Destim} we find that the scattering length $\beta$ can increase exponentially for an attractive scatterer. For quite small $r_0$ we have $k\beta\gg 1$. Thus we conclude that exponential increase of the scattering length for an attractive scatterer leads to the new characteristic length different from the radius of the scatterer.

From \eref{cirspec} we find that the cases of small ``metallic'' scatterers ($k\beta\ll 1$) and small attractive (``dielectric'') scatterers ($k\beta\gg 1$) lead to different signs of the shift of resonances of the resonator with respect to the empty one. For metallic scatterers resonances are shifted to higher wavenumbers and for the dielectric ones they are shifted to lower wavenumbers (see also \cite{stoe99}), which will become evident from the next section.

\section{\label{sec4}Wavefunction measurements}

In this section we consider the technique of wavefunctions measurements in the closed cavity. If the cavity is closed, its spectrum is discrete. The presence of antennas opens the cavity, thus all eigenvalues become resonances of the $S$-matrix. If some resonances are isolated from the rest, they can be considered separately and the influence of all others can be neglected. The same consideration takes place for an open cavity with a number of sharp resonances, isolated from the rest. The wavefunction corresponding to an isolated sharp resonance can be measured experimentally. Below we consider the idea of wavefunction measurements and some effects appearing if resonances are overlapping.

\begin{figure}
\includegraphics[width=7cm]{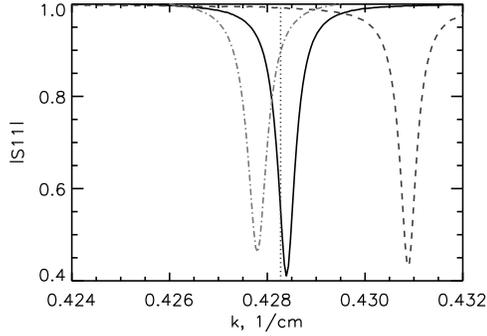}
\caption{\label{CavScatExpSpec}Experimentally measured resonances for the rectangular cavity ($47.3\times 20.0$~cm) with metallic and teflon scatterers. The vertical dotted line corresponds to the eigenwavenumber of the rectangle. The solid curve shows the modulus of the reflection coefficient from the empty cavity with a single exciting antenna. The dashed and the dashed-dotted curves show the resonances in the presence of a cylindrical metallic ($r_0=1.50$~mm) and dielectric (teflon, $r_0=2.56$~mm) scatterers positioned at $x_1=33.4$~mm, $x_2=50$~mm, respectively.}
\end{figure}

\subsection{One pole approximation}
Close to an eigenvalue $E_\mu$ of the discrete spectrum the Green function may be written as
\begin{equation}
G(\mathbf{r},\mathbf{R};k^2)\simeq\frac{\psi_\mu(\mathbf{r})\psi_\mu(\mathbf{R})}{E_\mu-k^2},
\label{G2parts}
\end{equation}
provided
$\mathbf{r}$ and $\mathbf{R}$ are far enough from each other. Function $\xi(\textbf{R};k^2)$ can be written in the form
\begin{eqnarray}
\xi(\textbf{R};k^2)=\frac{\psi^2_\mu(\mathbf{R})}{E_\mu-k^2}+\tilde\xi(\textbf{R};k^2),
\label{xi2parts}
\end{eqnarray}
where $\tilde\xi(\textbf{R};k^2)$ has no pole at $k^2=E_\mu$.
In one pole approximation \eref{xi2parts} reads
\begin{eqnarray}
\xi(\textbf{R};k^2)\simeq\frac{\psi^2_\mu(\mathbf{R})}{E_\mu-k^2}+\tilde\xi(\textbf{R};E_\mu).
\label{xi2parts1}
\end{eqnarray}
Using \eref{G2parts}, \eref{xi2parts1} one can calculate the inverse matrix appearing in Eq. \eref{Smatrix1} \cite{ste95}. Indeed in this case
\begin{equation}
\hat G+D-\frac{BC}{A+ik}=\alpha|\psi\rangle\langle\psi|+V,
\end{equation}
where $\alpha=(E_\mu-k^2)^{-1}$, $|\psi\rangle_i=\psi_\mu(\mathbf{R}_i)$,
$V=\widetilde D-BC/(A+ik)$,
$\widetilde D_{ij}=\widetilde D_i\delta_{ij}$ with the \textit{renormalized dimensionless} coefficient
\begin{equation}
\widetilde D_i=D_i+\tilde\xi(\textbf{R}_i;E_\mu).
\label{renD1ps}
\end{equation}
Using the identity
\begin{equation}
\frac{1}{\alpha|\psi\rangle\langle\psi|+V}=V^{-1}-
\frac{\alpha V^{-1}|\psi\rangle\langle\psi|V^{-1}}{1+\alpha\langle\psi|V^{-1}|\psi\rangle},
\end{equation}
we can rewrite Eq. \eref{Smatrix1} for $k^2$ close to $E_\mu$ in the form
\begin{eqnarray}
\fl S=-\frac{A-ik_\mu}{A+ik_\mu}+\frac{2ik_\mu B}{A+ik_\mu}
\left[V^{-1}-
\frac{V^{-1}|\psi\rangle\langle\psi|V^{-1}}{E_\mu-k^2+\langle\psi|V^{-1}|\psi\rangle}\right]
\frac{C}{A+ik_\mu},
\end{eqnarray}
where $k_\mu=\sqrt{E_\mu}$. This immediately gives us poles of the $S$-matrix:
\begin{equation}
k^2-E_\mu=\sum_{i=1}^\nu \frac{\psi^2_\mu(\mathbf{R}_i)}{\widetilde D_i-|B_i|^2/(A_i+ik_\mu)}.
\label{ShiftRes}
\end{equation}
Since nondiagonal elements $S_{ij},\,i\neq j$ are proportional to $\psi(\mathbf{R}_i)\psi(\mathbf{R}_j)$, they contain an additional information on the sign of the wavefunction.

Formula \eref{ShiftRes} \textit{implies the possibility to measure wavefunctions} through the shifts of resonances at different points of the cavity induced by a point like perturbation. In Ref.~\cite{sri91} the author used a movable point-like scatterer and fixed antenna whereas in \cite{ste95,ste92} the antenna has been moved.

Figure \ref{CavScatExpSpec} illustrates shifts of the resonance of the empty rectangular cavity (solid curve) as well as in the presence of a small metallic (dashed curve) and a dielectric (dashed-dotted curve). One sees that for a metallic scatterer the resonance is shifted to a higher wavenumber\footnote{For the larger metallic scatterer ($r_0=2.28$~mm) we observed a shift of the resonances to the lower wavenumbers in the low-$k$ regime ($k<0.9\,\textrm{cm}^{-1}$). This might be related to the fact that in the lower wavenumbers regime the distance between antenna and scatterer becomes comparable to the wavelength and the scattering length of the antenna. Then the proper description of the situation requires to go beyond one-pole approximation or even beyond the concept of a point-like scatterer. This experiment has to be investigated further in details.} while for the dielectric scatterer it is shifted to a lower wavenumber. This observation is in agreement with the arguments of the Section \ref{scatlest}.

\subsection{$m$-poles approximation}
In $m$-poles approximation one takes into account only contributions from $m$ poles in some range of wavenumbers. For $\mathbf{r}\neq \mathbf{R}$ we get an approximate expression for the Green function truncating the convergent series at the $m$th term:
\begin{equation}
G(\mathbf{r},\mathbf{R};k^2)\simeq\sum_{\mu=1}^m\frac{\psi_\mu(\mathbf{r})\psi_\mu(\mathbf{R})}{E_\mu-k^2}.
\label{npag}
\end{equation}
$\xi(\textbf{R};k^2)$ can be written in the form
\begin{eqnarray}
\xi(\textbf{R};k^2)=\sum_{\mu=1}^m\frac{\psi^2_\mu(\mathbf{R})}{E_\mu-k^2}+\tilde\xi(\textbf{R};k^2),
\end{eqnarray}
where $\tilde\xi(\textbf{R};k^2)$ has no poles for $E_1\geq k^2\leq E_m$. The approximation assumes that the energy dependence of $\tilde\xi(\textbf{R};k^2)$ can be neglected. Then we get approximately
\begin{eqnarray}
\xi(\textbf{R};k^2)\simeq\sum_{\mu=1}^m\frac{\psi^2_\mu(\mathbf{R})}{E_\mu-k^2}+\tilde\xi(\textbf{R};E),
\label{npaxi}
\end{eqnarray}
where $E$ is some value from the range $E_1\leq E\leq E_m$. Using \eref{npag}, \eref{npaxi} we can write
\begin{equation}
\hat G+D=W^+(H-k^2)^{-1}W+\widetilde D.
\end{equation}
Here $H_{\mu\mu'}=E_\mu \delta_{\mu\mu'}$, $W_{\mu i}=\psi_\mu(\textbf{R}_i)$,
$\widetilde D_{ij}=\widetilde D_i\delta_{ij}$, $\widetilde D_i=D_i+\tilde\xi(\textbf{R}_i;E)$. We see that inside of the given window of wavenumbers $\tilde\xi(\textbf{R}_i;E)$ is an offset renormalizing the scattering strength.

Let us write $S$-matrix \eref{Smatrix1} in $m$-poles approximation. We neglect energy dependence everywhere apart from the Green function. We have
\begin{equation}
\left[\hat G+D-\frac{BC}{A+ik_E}\right]^{-1}=\frac{1}{W^+(H-k^2)^{-1}W+V},
\end{equation}
where $k_E=\sqrt{E}$, $V=\widetilde D-BC/(A+ik_E)$. We write
\begin{eqnarray}
\frac{1}{V+W^+(H-k^2)^{-1}W}=\\=V^{-1}-V^{-1}W^+\frac{1}{H-k^2+WV^{-1}W^+}WV^{-1}.
\end{eqnarray}
Finally we get
\begin{eqnarray}
S=-\frac{A-ik_E}{A+ik_E}+2ik_E\frac{BCV^{-1}}{(A+ik_E)^2}-\nonumber\\-
\frac{2ik_EBV^{-1}}{A+ik_E}
W^+\frac{1}{H-k^2+WV^{-1}W^+}W\frac{V^{-1}C}{A+ik_E},\\
V^{-1}=\frac{\widetilde D(A^2+k_E^2)-BC(A+ik_E)}{(A\widetilde D-BC)^2+k_E^2\widetilde D^2}=V^{-1}_R-iV^{-1}_I.
\end{eqnarray}
If $\widetilde D\gg 1$ then
\begin{eqnarray}
V^{-1}_R\simeq \widetilde D^{-1}, \quad V^{-1}_I\simeq \frac{kBC}{\widetilde D^2(A^2+k_E^2)}.
\end{eqnarray}
Introducing an effective Hamiltonian
\begin{equation}
H_{\scriptsize{\textrm{eff}}}=H+WV^{-1}_RW^+ - iWV^{-1}_IW^+,
\label{Heff2}
\end{equation}
we can write
\begin{eqnarray}
\fl S=-\frac{A-ik_E}{A+ik_E}+2ik_E\frac{BCV^{-1}}{(A+ik_E)^2}-
\frac{2ik_EBV^{-1}}{A+ik_E}
W^+\frac{1}{H_{\scriptsize{\textrm{eff}}}-k^2}W\frac{V^{-1}C}{A+ik_E}.
\label{Smatrix5}
\end{eqnarray}
Experimentally one usually removes the global phase of $S$-matrix, describing the antenna matching itself. Theoretically this means that instead of \eref{Smatrix5} one introduces a new matrix $S'$
\begin{eqnarray}
\fl S'=-\frac{A+ik_E}{A-ik_E}S=1-2ik_E\frac{BCV^{-1}}{A^2+k_E^2}+
\frac{2ik_EBV^{-1}}{A-ik_E}
W^+\frac{1}{H_{\scriptsize{\textrm{eff}}}-k^2}W\frac{V^{-1}C}{A+ik_E}.
\end{eqnarray}
Poles of $S$-matrix can be found from the equation
\begin{equation}
\det[H_{\scriptsize{\textrm{eff}}}-k^2]=0.
\label{poleHeff}
\end{equation}
For the closed system the last equation reduces to
\begin{equation}
\det[H+W\widetilde D^{-1}W^+-k^2]=0.
\end{equation}
Effective Hamiltonians of the type \eref{Heff2} were used in \cite{stoe99,mah69,lew91}. Resonances of $S$-matrix in a general situation are discussed in \cite{rot01}.

\subsection{Experimental ``wavefunctions'' in the case of degenerated states}
Let us consider a cavity with the degenerated eigenvalue $E_\mu=E,\,\mu=1,\ldots,m$. In this case any weak perturbation mixes all eigenstates corresponding to $E$. In the wavefunction measurement an antenna is a perturbation. Moving the antenna one changes mixing coefficients and produces a new field configuration inside the cavity. Therefore it is not clear whether one can easily treat the experimental wavefunction in this case. We show below that the experimental pattern in such a situation is just a sum of squares of all eigenfunctions $\psi_\mu(\mathbf{r})$. This combination is invariant with respect to the choice of eigenfunctions.

The simplest approximation in this situation is to take into account only $m$ eigenfunctions corresponding to $E$. Let us calculate an experimental pattern which can be reconstructed from the measurement with a single antenna. The resonance position for the fixed antenna position $\textbf{R}$ can be calculated from Eq.~\eref{poleHeff}. Using Eq.~\eref{npaxi} we get
\begin{eqnarray}
\sum_{\mu=1}^m\frac{\psi^2_\mu(\mathbf{R})}{E_\mu-k^2}+V
=\sum_{\mu=1}^m\frac{\psi^2_\mu(\mathbf{R})}{E-k^2}+V=0,\\
k^2-E=V^{-1}\sum_{\mu=1}^m \psi^2_\mu(\mathbf{R}).
\label{degShift}
\end{eqnarray}
From Eq.~\eref{degShift} we conclude that in the case of degenerated eigenvalue the experimentally measured pattern should correspond to the sum of squares of corresponding eigenfunctions.

\begin{figure}
\parbox{4.2cm}{
\parbox{.4cm}{a)}
\parbox{3.7cm}{\includegraphics[width=3.7cm]{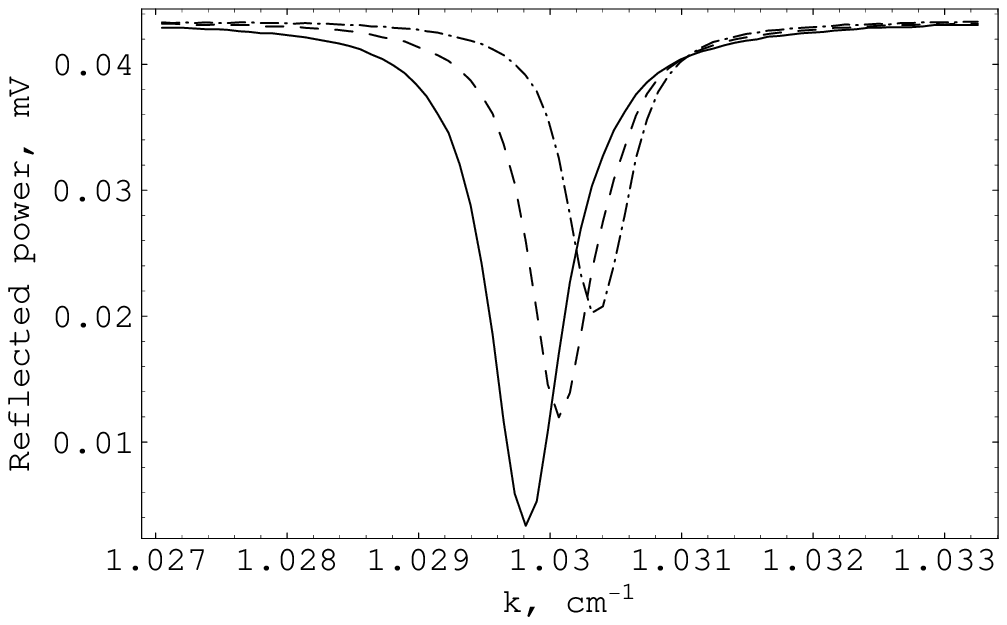}}
\parbox{.4cm}{c)}
\parbox{3.7cm}{\includegraphics[width=3.7cm]{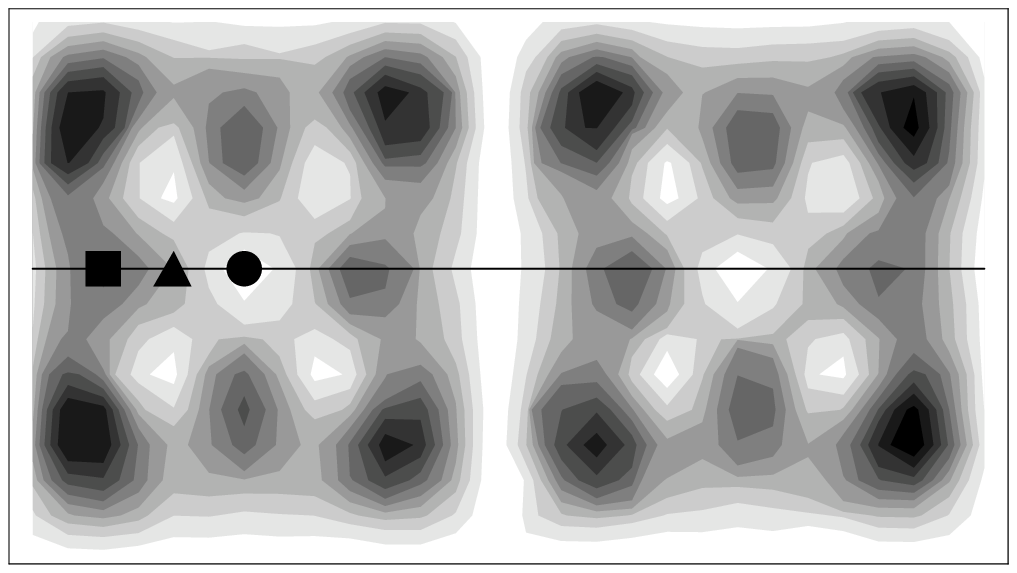}}
\parbox{.4cm}{e)}
\parbox{3.7cm}{\includegraphics[width=3.7cm]{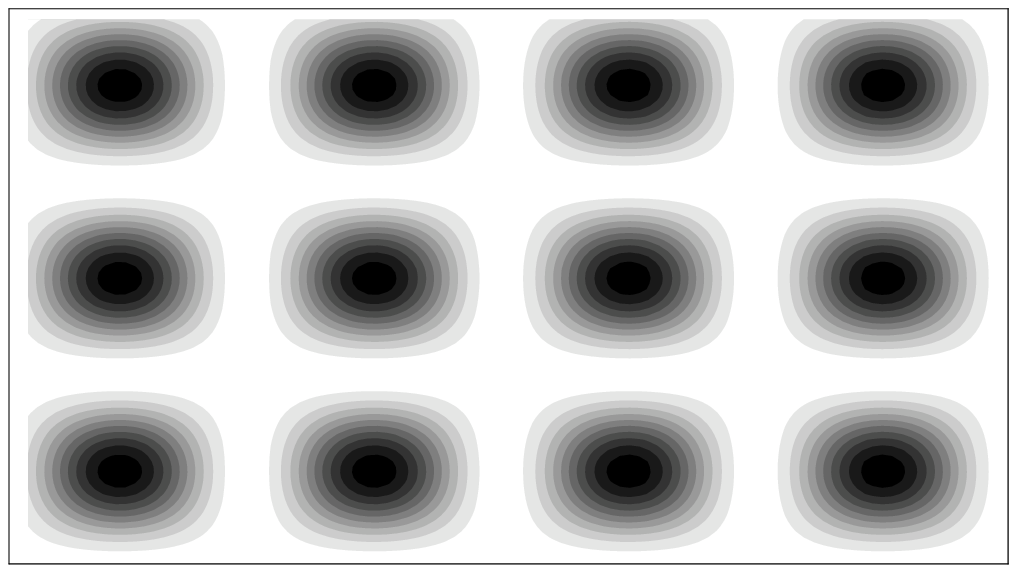}}
\parbox{.4cm}{g)}
\parbox{3.7cm}{\includegraphics[width=3.7cm]{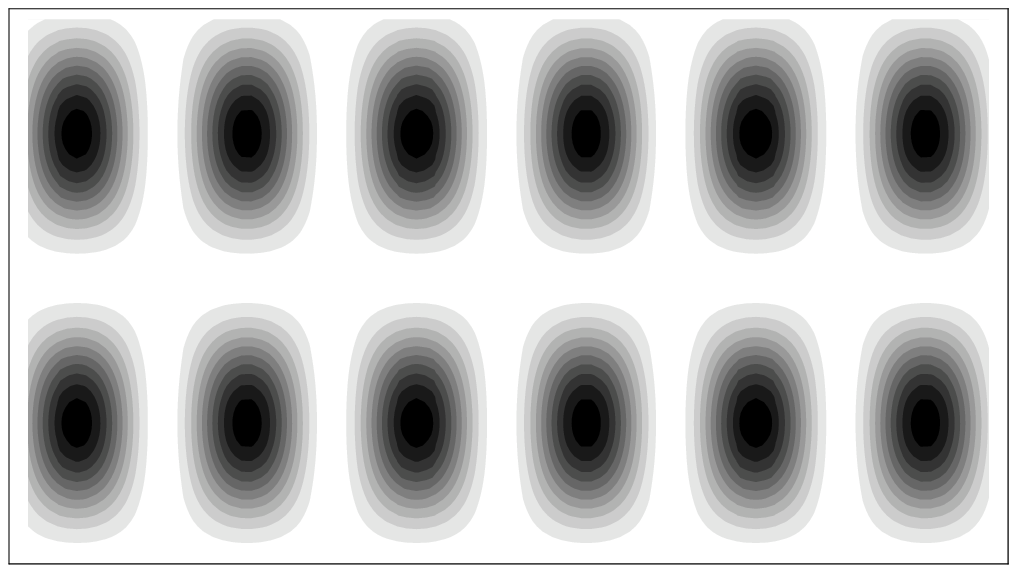}}
\parbox{.4cm}{i)}
\parbox{3.7cm}{\includegraphics[width=3.7cm]{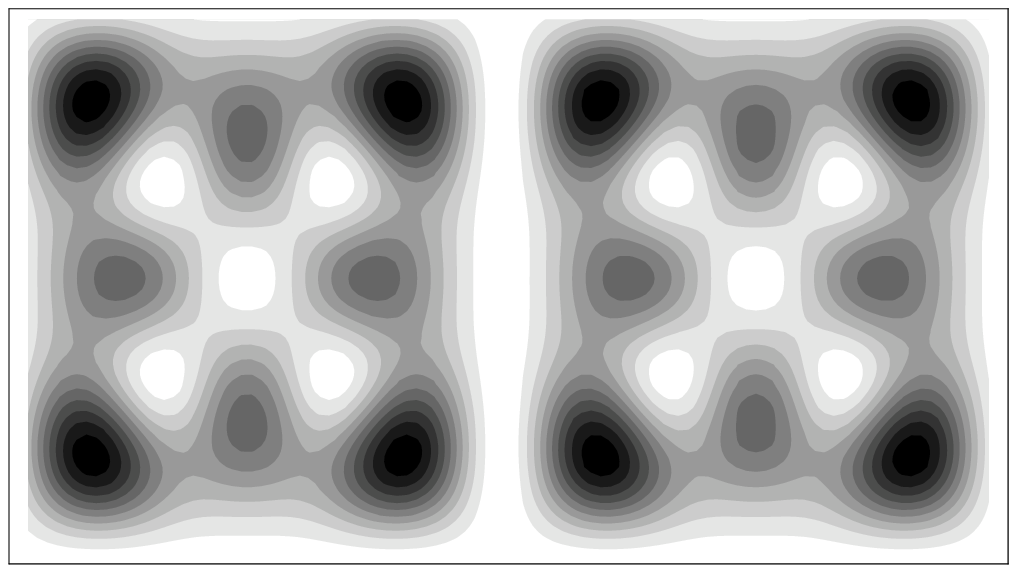}}
}
\parbox{4.2cm}{
\parbox{.4cm}{b)}
\parbox{3.7cm}{\includegraphics[width=3.7cm]{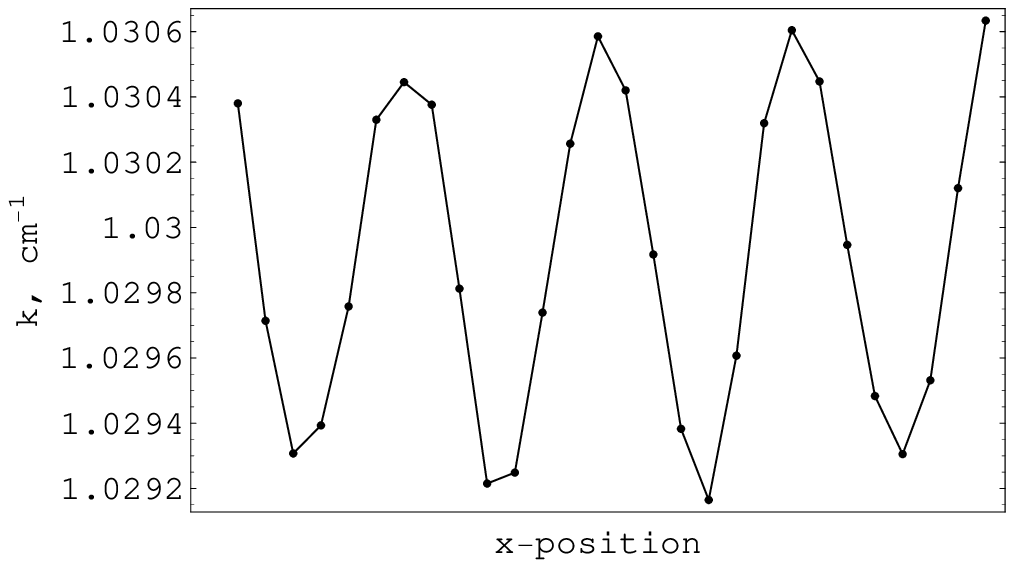}}\\
\parbox{.4cm}{d)}
\parbox{3.7cm}{\includegraphics[width=3.7cm]{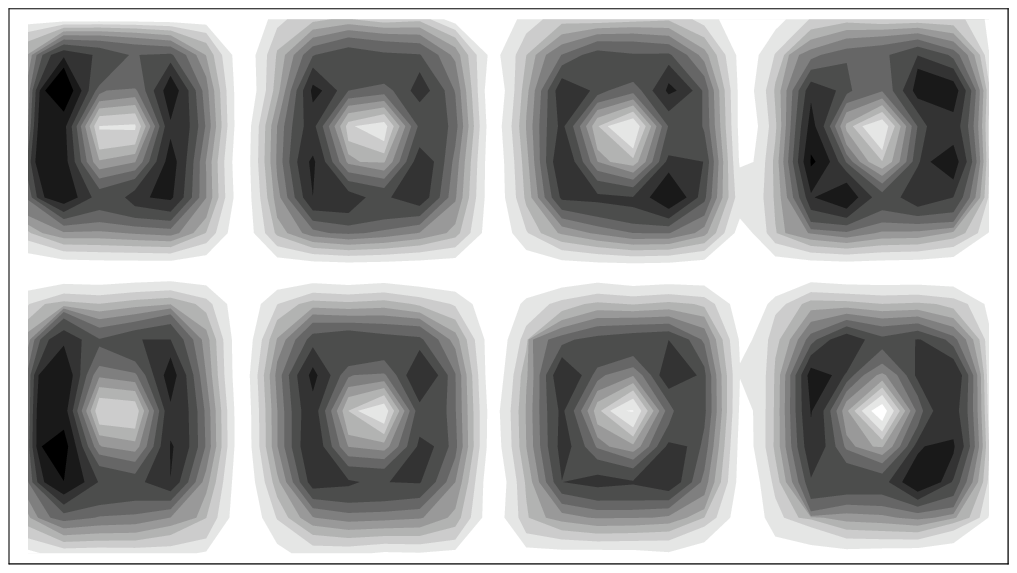}}
\parbox{.4cm}{f)}
\parbox{3.7cm}{\includegraphics[width=3.7cm]{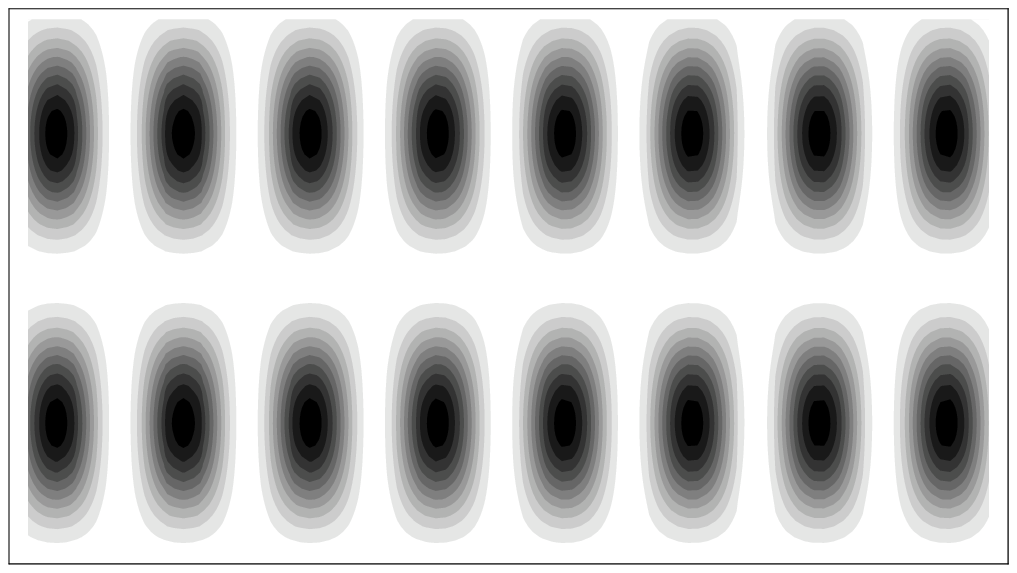}}\\
\parbox{.4cm}{h)}
\parbox{3.7cm}{\includegraphics[width=3.7cm]{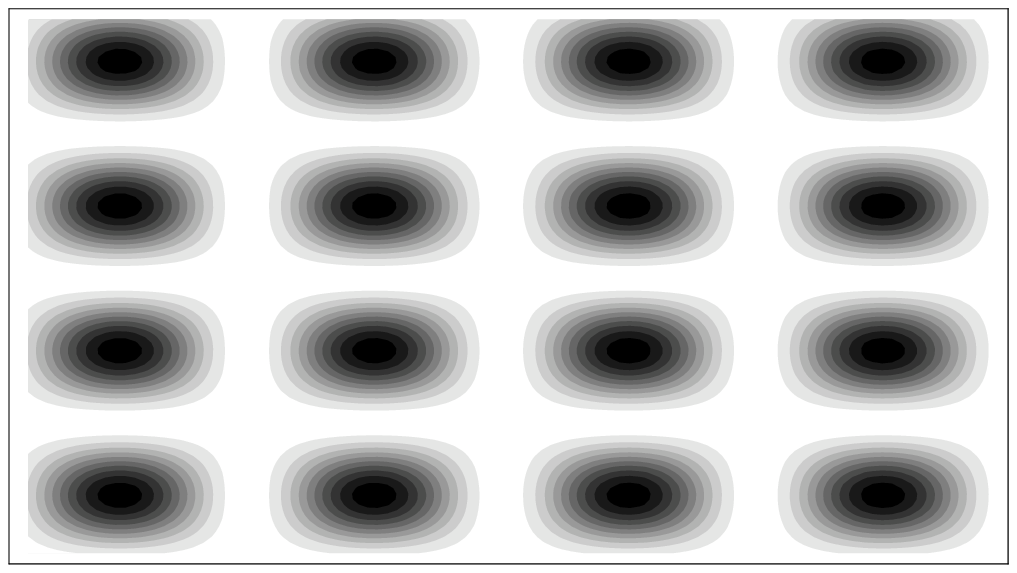}}\\
\parbox{.4cm}{j)}
\parbox{3.7cm}{\includegraphics[width=3.7cm]{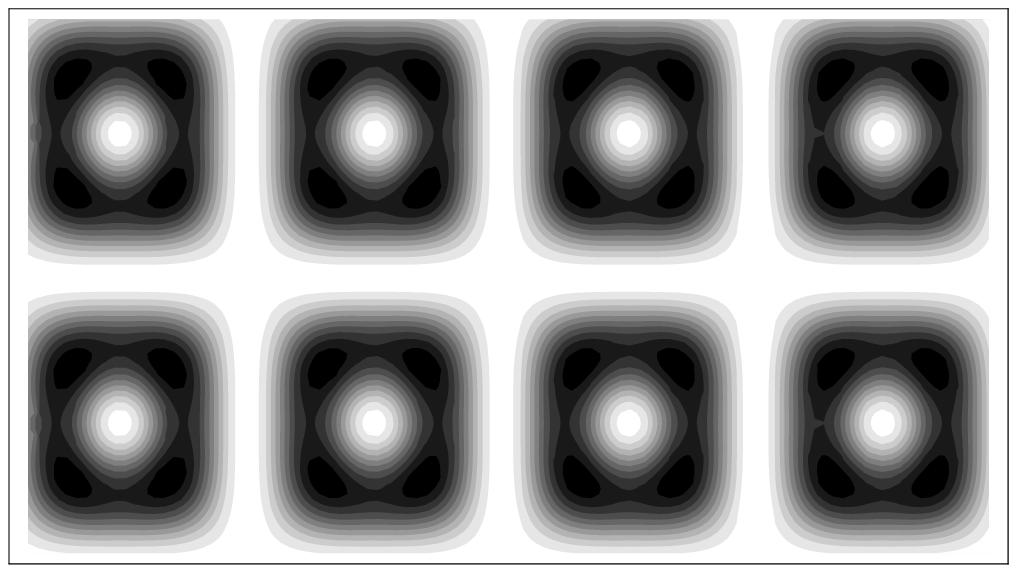}}
}
\caption{Comparison of experiment and theory for two degenerated states of the rectangular billiard with side ratio $2:1$: a) experimental resonances for the degenerated $15$th and $16$th states for three antenna positions marked by a square, a triangle and a disk in c); b) the resonance wavenumber corresponding to $15$th, $16$th states versus antenna position on the horizontal line in c); c) resonance wavenumber-shift pattern in a gray-scale; e) $|\psi_1|^2$; g) $|\psi_2|^2$; i) $|\psi_1|^2+|\psi_2|^2$ for the $15$th and $16$th states; d), f), h), j) the same for the degenerated $25$th and $26$th states}
\label{DegStateRect}
\end{figure}

In Fig.~\ref{DegStateRect} experimental examples from a rectangular microwave billiard with a side ratio of 2:1 are presented. For the technical details the reader is referred to Ref.~\cite{kuh00b}. The figure shows experimental and theoretical patterns corresponding to doublets consisting of the $15$th, $16$th states and the $25$th, $26$th states. States of each doublet are degenerated, the data were obtained from the reflection measurement with a single movable antenna. Thus one expects to measure the sum of squares of two eigenfunctions. Pictures a)-d) illustrate experimental data, all others show theoretical simulations.

Fig.~\ref{DegStateRect}a) shows the resonances corresponding to the $15$th state with the antenna at the positions marked by a square, a triangle and a disk at Fig.~\ref{DegStateRect}c). Solid, dashed and dashed-dotted lines in Fig.~\ref{DegStateRect}a) correspond to a square, a triangle and a disk, respectively. Here only a single resonance is observed and no splitting could be seen. Fig.~\ref{DegStateRect}b) shows the shift of the resonance while the antenna is shifted along the horizontal line in Fig.~\ref{DegStateRect}c). In Figs.~\ref{DegStateRect}c) and d) the obtained resonance wavenumber shifts patterns are plotted in a gray-scale. In Figs.~\ref{DegStateRect}e)-h) the squares $|\psi_1|^2$, $|\psi_2|^2$ of the rectangular eigenstates are plotted. In Figs.~\ref{DegStateRect}i)-j) $|\psi_1|^2+|\psi_2|^2$ is plotted, demonstrating the perfect agreement with the experimental results shown in Figs.~\ref{DegStateRect}c)-d).

\section{\label{sec5}Transmission measurements in open systems}

Recent measurements of nodal domains in open microwave cavities \cite{kuh07a,kuh07b} raised a question on the theoretical description of this type of experiments. The crucial difference compared to the case considered above  of a closed cavity consists in the absence of sharp resonances of the $S$-matrix. This makes it impossible to use the one-pole approximation being the base of wavefunctions measurements in closed systems.

In the experiment the transmission coefficient from the fixed antenna to the movable one is measured and interpreted directly as a ``wavefunction'' corresponding to the continuous spectrum of the unperturbed open system. We show below that in fact in the experiment the Green function of the cavity is measured. To prove this statement we start from Eq.~\eref{Smatrix1}. The straightforward calculation gives for the case of two antennas
\begin{eqnarray}
S_{12}(\mathbf{R}_1,\mathbf{R}_2;k^2)=
-\frac{B_1C_2}{\lambda(A_1+ik)(A_2+ik)}G(\mathbf{R}_1,\mathbf{R}_2;k^2),\label{s12}\\
\lambda=\lambda(\mathbf{R}_1,\mathbf{R}_2;k^2)=
\det\left[\hat G(\mathbf{R}_1,\mathbf{R}_2;k^2)+D-\frac{BC}{A+ik}\right].
\label{lambda}
\end{eqnarray}
Let us assume that $k$ is far from all poles $k_\mu$ and scattering is weak, i.e.
\begin{equation}
\widetilde D_i\equiv D_i-(2\pi)^{-1}[\ln(k/2)+\gamma]=
-(2\pi)^{-1}[\ln(k\beta_i/2)+\gamma]\to\infty.
\end{equation}
Here we used \eref{scldef}. Thus for quite distant antennas we can write
\begin{equation}
\lambda\simeq \widetilde D_1\widetilde D_2 \ldots \widetilde D_\nu.
\label{lamas}
\end{equation}
Measuring $S_{12}(\mathbf{R}_1,\mathbf{R}_2;k^2)$ at a fixed wavenumber and a fixed antenna position $\mathbf{R}_2$ we get a value proportional to $G(\mathbf{R}_1,\mathbf{R}_2;k^2)$. An additional phase shift in $S_{12}$ due to the prefactor of the Green function in Eq.~\eref{s12} can be easily removed for the closed cavity, since the Green function in this case is real.

\subsection{Example: Single antenna attached to a semi-infinite cavity}

The following example illustrates general effects taking place in the vicinity of a boundary. We consider two antennas attached to a semi-infinite cavity at points $\mathbf{R}_1,\,\mathbf{R}_2$. The cross-section of the cavity is a half plane.

Let the boundary of the cavity coincides with $x_2$-axis in the plane. We define  $\mathbf{R}_1=(x_1,x_2),\,\mathbf{R}_2=(\tilde x_1,\tilde x_2)$.
The transmission coefficient $S_{12}$ through these antennas is defined by \eref{s12} where $G(\mathbf{R}_1,\mathbf{R}_2;k^2),\,\xi(\mathbf{R}_i;k^2)$ can be found from the method of images:
\begin{equation}
\fl
G(\mathbf{R}_1,\mathbf{R}_2;k^2)=\frac{i}{4}
\Bigl[H_0^{(1)}(k|\mathbf{R}_1-\mathbf{R}_2|)-H_0^{(1)}(k|\mathbf{R}_1-\mathbf{R}_2'|)\Bigr], \quad
\mathbf{R}_2'=(-\tilde x_1,\tilde x_2),
\end{equation}
\begin{equation}
\xi(\mathbf{R}_1;k^2)=-\frac{1}{2\pi}\Bigl(\ln\frac{k}{2}+\gamma\Bigr)+
\frac{i}{4}\Bigl(1-H_0^{(1)}(2kx_1)\Bigr).
\end{equation}
For simplicity we assume $k|\mathbf{R}_1-\mathbf{R}_2|\gg 1$ and neglect nondiagonal elements in $\hat G$ in \eref{lambda}. In this approximation
\begin{equation}
\lambda=\left(\xi(\mathbf{R}_1;k^2)+D_1-\frac{B_1C_1}{A_1+ik}\right)\cdot
\left(\xi(\mathbf{R}_2;k^2)+D_2-\frac{B_2C_2}{A_2+ik}\right).
\end{equation}
If the movable antenna is placed far from the boundary then $kx_1\gg 1$. Using the asymptotic expansion of the Hankel function \cite{gra80} and \eref{scldef} we write
\begin{equation}
\xi(\mathbf{R}_1;k^2)+D_1\simeq-\frac{1}{2\pi}\ln(k\beta_1)+
\frac{i}{4}\Bigl(1-\frac{e^{i(2kx_1-\pi/4)}}{\sqrt{\pi k x_1}}\Bigr).
\end{equation}
Thus we see that at distances of the order of the wavelength from the boundary $\lambda$ contains an oscillating term. This term decays when one moves the antenna away from the boundary. The origin of this oscillating part is the mirror source of the movable antenna. This mirror source can be treated as an ``impurity'' charge, thus the oscillating behavior of $\lambda$ may be interpreted in terms of Friedel oscillations \cite{bac08}.

Let us now approach the movable antenna to the boundary.
If $kx_1\ll 1$ then using \eref{G2Dplane}, \eref{scldef} we write
\begin{equation}
\xi(\mathbf{R}_1;k^2)+D_1\simeq \frac{1}{2\pi}\left(\ln\frac{x_1}{\beta_1}+\gamma\right).
\end{equation}
The comparison of the last formula with an assumption \eref{lamas} shows that at distances less than the wavelength the measured quantity can differ from the Green function.

\section{\label{sec6}Cavity with closed loops}

\begin{figure}
\includegraphics[width=7cm]{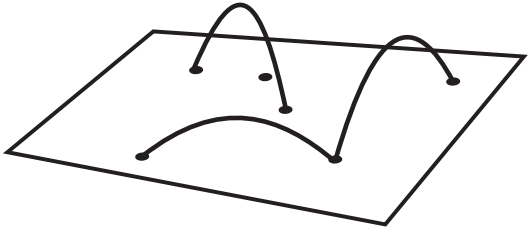}
\caption{\label{CavLoops}Cavity with point-like scatterers and closed loops.}
\end{figure}

In this section we consider a cavity with closed loops (see Figure \ref{CavLoops}). This model is another realization of a quantum graph \cite{kot03}. In order to simplify calculations we do not consider attached antennas.

\begin{figure}
\includegraphics[width=7cm]{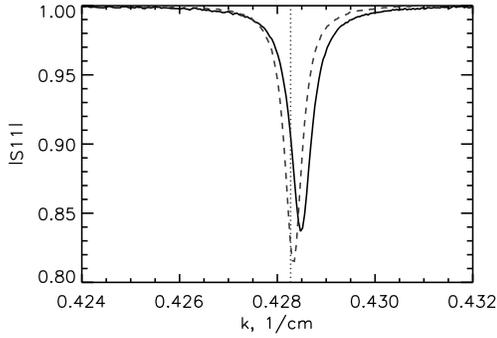}
\caption{\label{CavLoopsExpSpec}Shift of the same resonance of the rectangular cavity as shown in Fig.~\ref{CavScatExpSpec} due to a loop connecting two antenna at positions $(14.0,8.0)$~cm, $(40.4,5.0)$~cm. Solid line: both antennas are closed by 50~$\Omega$ load at the end, dashed line: the antennas are connected by a wire. Dotted line: eigenwavenumber of the rectangle.
}
\end{figure}

\subsection{Cavity with short loops}
Analogously to Eq.~\eref{OperDef} we introduce the Hamiltonian inside the cavity
\begin{equation}
H\varphi=-\Delta\varphi.
\label{OperClosedCav}
\end{equation}
Repeating steps \eref{rhoforms}-\eref{formsdifference} we get
\begin{equation}
\left<\varphi,H\psi\right> - \overline{\left<\psi,H\varphi\right>}=
\left<\eta_\varphi,\zeta_\psi\right>-\left<\zeta_\varphi,\eta_\psi\right>,
\end{equation}
where $\eta_\varphi,\,\zeta_\varphi,\,\eta_\psi,\,\zeta_\psi$ are vectors with components $\eta_\varphi^{i}={L_{1i}^{\varphi_2}},\,\zeta_\varphi^i=L_{0i}^{\varphi_2}$, $\eta_\psi^{i}=L_{1i}^{\psi_2},\,\zeta_\psi=L_{0i}^{\psi_2}$.

Analogously to \eref{CouplConst} we write
\begin{equation}
\eta_\varphi=D \zeta_\varphi,\quad \eta_\psi=D\zeta_\psi.
\label{loops-matching}
\end{equation}
Here $D$ is a Hermitian matrix with elements $D_{ij}$, $i,\,j=1,\ldots,\nu$, where $\nu$ is the number of inserts including both scatterers and points of loops attachments. Elements $D_{ij}=0,\,i\neq j$ if points $i$ and $j$ are not connected. In contrast to \eref{Smatrix1} matrix $D$ contains non-diagonal elements describing loops.

Using matching condition \eref{loops-matching} we can find eigenwavenumbers of the closed cavity. We look for the solution in the form
\begin{equation}
\varphi(\mathbf{r})=\sum_{i=1}^\nu F_i G(\mathbf{r},\mathbf{R}_i;k^2).
\end{equation}
Substituting \eref{L0L1} in \eref{loops-matching} we get
\begin{equation}
[\hat G(\mathbf{R}_1,\ldots,\mathbf{R}_\nu;k^2) + D]F=0,
\end{equation}
where $F=(F_1,\ldots,F_\nu)$. Thus the equation for the eigenvalues $k^2$ is
\begin{equation}
\det[\hat G(\mathbf{R}_1,\ldots,\mathbf{R}_\nu;k^2) + D]=0.
\end{equation}

\subsection{A more realistic model of the loop}

The model of loops considered above is appropriate only for short loops or for some narrow energy region since it does not contain the spectrum of the one-dimensional wire connecting different points. To take into account the ``inner structure'' of the loop let us consider a more realistic model. We consider a cavity with a loop of the length $L$ matched to it at two points $\mathbf{R}_1,\,\mathbf{R}_2$. We neglect the curvature of the wire. Analogously to Eq.~\eref{ManyAntennasStat} we write
\begin{equation}
\begin{array}{l}
\varphi_{1}(x)=a e^{ikx}+b e^{-ikx},\\
\varphi_2(\mathbf{r})=\sum_{i=1}^2 F_i G(\mathbf{r},\mathbf{R}_i;k^2).
\end{array}
\end{equation}
Matching conditions \eref{ant-matching} at points $\mathbf{R}_1,\,\mathbf{R}_2$ give
\begin{eqnarray}
\fl
\left(\begin{array}{l} ik(a-b) \\ F_1\xi(\mathbf{R}_1;k^2)+
F_2 G(\mathbf{R}_1,\mathbf{R}_2;k^2)
\end{array}\right)
=\left(\begin{array}{ll} A_1 & B_1 \\ C_1 & D_1 \end{array}\right)
\left(\begin{array}{l}
a_1 + b_1 \\ -F_1
\end{array}\right),\nonumber\\
\fl
\left(\begin{array}{l} ik(a e^{ikL} -b e^{-ikL}) \\ F_2\xi(\mathbf{R}_2;k^2)+
F_1 G(\mathbf{R}_2,\mathbf{R}_1;k^2)
\end{array}\right)
=\left(\begin{array}{ll} A_2 & B_2 \\ C_2 & D_2 \end{array}\right)
\left(\begin{array}{l}
a e^{ikL} + b e^{-ikL} \\ -F_2
\end{array}\right).
\label{LoopMatch2points}
\end{eqnarray}
From Eq.~\eref{LoopMatch2points} we find that the spectrum of the system is defined by the equation
\begin{equation}
\det\left(\begin{array}{ll}
Z_1 & B \\
Z_2 & \hat G + D
\end{array}\right)=0,
\label{LoopSpec}
\end{equation}
where $Z_1,\,Z_2$ are $2\times 2$ matrices
\begin{eqnarray}
Z_1=
\left(\begin{array}{ll}
A_1-ik & A_1+ik \\
(A_2-ik)e^{ikL} & (A_2+ik)e^{-ikL}
\end{array}\right),\\
Z_2=\left(\begin{array}{ll}
C_1 & C_1 \\
C_2e^{ikL} & C_2e^{-ikL}
\end{array}\right).
\end{eqnarray}
If the cavity and the wire are decoupled then $B=0,\,Z_2=0$ and Eq.~\eref{LoopSpec} reduces to
\begin{equation}
\det(Z_1)\det(\hat G + D)=0.
\end{equation}
The last equation gives the unification of spectra of the cavity with scatterers inside and the wire.

In Fig.~\ref{CavLoopsExpSpec} one sees the experimentally observed shift of the same resonance of the rectangular cavity as shown in Fig.~\ref{CavScatExpSpec}. The dotted line shows the position of the eigenwavenumber of the rectangular cavity. The solid line corresponds to the modulus of the reflection coefficient from the cavity excited by the antenna with two additionally attached antennas each closed by a 50~$\Omega$ load. The dashed line illustrates the modulus of the reflection coefficient from the same cavity, were the antennas were connected by a wire.

\section{\label{sec7}The $S$-matrix for multimode cavity}

\begin{figure}
\includegraphics[width=5cm]{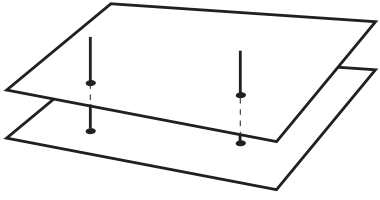}
\caption{\label{NChannels}The illustration of the graph corresponding to 2 open transversal modes.}
\end{figure}

If there are more than one open transversal modes in the cavity with transversal wavenumbers $k_\perp^{(n)}<k,\,n=1,\ldots,N$ the ansatz of the solution \eref{ManyAntennasStat} has to be generalized. To treat this situation
let us consider $N$ identical ``copies'' of the cavity (see figure \ref{NChannels}) with wavefunctions $\varphi_{2}(\mathbf{r}),\ldots\varphi_{N+1}(\mathbf{r})$ and introduce the operator acting on the spacial variables of the antenna and the $N$ cavities:
\begin{equation}
H\varphi=\left(\begin{array}{l}-\varphi_1''(x) \\ (-\Delta+(k_\perp^{(1)})^2)\varphi_2(\mathbf{r}) \\
\vdots \\ (-\Delta+(k_\perp^{(N)})^2)\varphi_{N+1}(\mathbf{r})\end{array}\right),
\label{OperDefN}
\end{equation}
where $k_\perp^{(n)}$ is $n$-th transversal wavenumber.
We look for the solution of the spectral problem \eref{SpecProb} in the form
\begin{eqnarray}
\varphi_{1i}(x)=a_i e^{ikx}+b_i e^{-ikx}, \nonumber\\
\varphi_{n+1}(\mathbf{r})=\sum_i F_i^{(n)} G(\mathbf{r},\mathbf{R}_i;(k_\parallel^{(n)})^2),
\end{eqnarray}
where $k_\parallel^{(n)}=[k^2-(k_\perp^{(n)})^2]^{1/2},\,n=1,\ldots,N$ (for the lowest TM-mode in electromagnetic billiards $k_\perp^{(1)}=0$ and $k_\parallel^{(1)}=k$).

The same procedure as for a single transversal channel gives us the following expression for the $S$-matrix:
\begin{eqnarray}
S=-\frac{M-ik}{M+ik},
\end{eqnarray}
where $M=A-Z$, $A_{ij}=A_i\delta_{ij}$,
\begin{equation}
Z_{ij}=\sum_{nn'}B^{(n)}_i [Q^{-1}]^{(nn')}_{ij}C^{(n')}_{j}.
\end{equation}
Here the matrix $Q$ has elements
\begin{equation}
Q^{(nn')}_{ij} =\hat G_{ij}^{(n)}\delta_{nn'}+D_i^{(nn')}\delta_{ij},
\end{equation}
where $\hat G_{ij}^{(n)}=\hat G_{ij}(\mathbf{R}_i,\mathbf{R}_j;(k_\parallel^{(n)})^2)$.

\section{Conclusions}

In the paper we presented the general selfconsistent theory for measurements on electromagnetic cavities with point-like antennas and scatterers. In the framework of the considered approach we justified the idea lying on the heart of wavefunctions measurement. The experimental patterns for the cases of degenerated states are explained. The generalization of the basic idea allowed to describe cavities with more than one open transversal channel and antennas connected by loops. Mathematical objects appearing in the last cases can be considered as quantum graphs.

\section{Acknowledgements}

We gratefully acknowledge useful discussions with Petr $\check{\textrm{S}}$eba. This work was supported by the Deutsche Forschungsgemeinschaft via an individual grant.

\appendix
\section{\label{appA}Operator algebra for point scatterers in a closed cavity.}
Similar to the one-dimensional case one usually treats a point-like scatterer as a $\delta$-potential which can be written in an arbitrary representation as a matrix $WW^+$, where $W$ is a coupling matrix \cite{stoe99}. However in a multidimensional case a point-like perturbation does not correspond exactly to a $\delta$-potential. We show below that instead of a $\delta$-potential in the two-dimensional case a point-like scatterer situated at the point $\textbf{R}$ is described by a product $\delta(\textbf{r}-\textbf{R})\mathcal{R}$, where $\mathcal{R}$ is a \textit{renormalizing operator}.

From Eqs.~\eref{assol}, \eref{L0L1} we find that the function $\varphi_2$ obeys the equation
\begin{equation}
(H - k^2)\varphi_2=\sum_{i=1}^\nu F_i \delta(\mathbf{r}-\mathbf{R}_i).
\end{equation}
Using Eq.~\eref{ant-matching} with $B=C=0$ we write
\begin{equation}
(H - k^2)\varphi_2(\mathbf{r})=-\sum_{i=1}^\nu D_i^{-1}\delta(\mathbf{r}-\mathbf{R}_i)\mathcal{R} \varphi_2(\mathbf{r})
\label{NScatRightDelta}
\end{equation}
where
\begin{eqnarray}
\mathcal{R}\varphi_2(\mathbf{r})=
\lim_{\mathbf{r'}\to\mathbf{r}}
[(1-\ln(|\mathbf{r}'-\mathbf{r}|)(\mathbf{r}'-\mathbf{r})\nabla)\varphi_2(\mathbf{r}')].
\end{eqnarray}
Thus $\mathcal{R}$ acts on the function $\varphi(\mathbf{r})$ without a singularity at the point $\mathbf{r}$ as an identical operator and gives the renormalized value for a function with a logarithmic singularity (see Eq.~\eref{assol}).

In the basis of eigenfunctions $\psi_\mu(\mathbf{r})$ the matrix elements of the $\delta$-function are given by
\begin{equation}
\langle\psi_{\mu}(\mathbf{r})|\delta(\mathbf{r}-\mathbf{R})|\psi_{\mu'}(\mathbf{r})\rangle=
\psi_{\mu}(\mathbf{R})\psi_{\mu'}(\mathbf{R}).
\end{equation}
In this basis Eq.~\eref{NScatRightDelta} reads
\begin{equation}
(H - k^2)|\varphi_2\rangle=-W D^{-1} W^+ \mathcal{R}|\varphi_2\rangle,
\label{NScatRightDelta2}
\end{equation}
where $|\varphi_2\rangle_\mu=\langle\psi_{\mu}|\varphi_2\rangle$, $W_{\mu i}=\psi_\mu(\textbf{R}_i)$.
The solvability condition of Eq.~\eref{NScatRightDelta} is 
\begin{equation}
\det\biggl[H - k^2+W D^{-1} W^+ \mathcal{R}\biggr]=0.
\end{equation}
The last equality can be rewritten in the form
\begin{eqnarray}
\det\biggl[H - k^2+W D^{-1} W^+ \mathcal{R}\biggr]=\det[H - k^2]\times\nonumber\\
\times\det\biggl[1+W D^{-1} W^+ \mathcal{R}(H - k^2)^{-1}\biggr]=\det[H - k^2]\times\nonumber\\
\times\det[D^{-1}]\det\biggl[D+W^+ \mathcal{R}(H - k^2)^{-1}W\biggr]=0.
\label{OperDet}
\end{eqnarray}

For any vectors $|\varphi\rangle,\,|\psi\rangle$ we have $\langle \psi|\mathcal{R}|\varphi\rangle=\langle \psi|\varphi\rangle$ if $\langle \psi|\varphi\rangle$ exists. Thus for $i\neq j$ we have
\begin{eqnarray}
[W^+ \mathcal{R}(H - k^2)^{-1}W]_{ij}=
\sum_{\mu}\frac{\psi_\mu(\textbf{R}_i)\psi_\mu(\textbf{R}_j)}{E_\mu - k^2}.
\end{eqnarray}
From this we conclude that
\begin{equation}
W^+ \mathcal{R}(H - k^2)^{-1}W=\hat G.
\end{equation}
Finally the equation for the spectrum reads
\begin{equation}
\det[H-k^2]\det[\hat G+D]=0.
\end{equation}
The last equation is in accordance with \eref{SpecScat}.

\section{\label{appB}Green function for the system with antennas.}

Let us consider a cavity with a number of fixed antennas. In order to measure the wavefunction of such a system one has to introduce a movable antenna. The wavefunction of the cavity \textit{with a movable antenna} is the Green function of the cavity \textit{with only fixed antennas}. Then the natural question is to calculate Green function of the cavity with a number of fixed antennas.

Let us look for the Green function of the system with fixed antennas $G_p$ in the form
\begin{equation}
G_p(\mathbf{r},\mathbf{R};k^2)=G(\mathbf{r},\mathbf{R};k^2)+
\sum_j F_j G(\mathbf{r},\mathbf{R}_j;k^2),
\label{Ganzatz}
\end{equation}
where $G(\mathbf{r},\mathbf{R};k^2)$ is the Green function for the cavity without antennas.
The function \eref{Ganzatz} satisfies the equation for the Green function. The Green function should correspond to outgoing waves spreading inside of the fixed antennas, thus inside of the $i$th fixed antenna the solution is
\begin{equation}
\varphi_{1i}(x)=b_i e^{-ikx}.
\end{equation}
In the neighborhood of $\mathbf{R}_i$ the Green function has the following asymptotic
\begin{eqnarray}
G_p(\mathbf{r}\to \mathbf{R}_i,\mathbf{R};k^2)\to G(\mathbf{R}_i,\mathbf{R};k^2)+\nonumber\\+
\sum_{j\neq i} F_j G(\mathbf{R}_i,\mathbf{R}_j;k^2)+
F_i\left[-\frac{1}{2\pi}\ln(|\mathbf{r}-\mathbf{R}_i|)+\xi(\mathbf{R}_i;k^2)\right].
\end{eqnarray}
Using the condition \eref{ant-matching} we get
\begin{eqnarray}
\fl
\left(\begin{array}{l}
-ikb_i \\
G(\mathbf{R}_i,\mathbf{R};k^2)+
\sum_{j\neq i} F_j \hat G(\mathbf{R}_i,\mathbf{R}_j;k^2)
\end{array}\right)=
\left(\begin{array}{ll}
A_i & B_i \\
C_i & D_i
\end{array}\right)
\left(\begin{array}{l}
b_i \\
-F_i
\end{array}\right).
\end{eqnarray}
From the last equality analogously to \eref{eq01}, \eref{eq02} we find
\begin{eqnarray}
\fl
G_p(\mathbf{r},\mathbf{R};k^2)=G(\mathbf{r},\mathbf{R};k^2)-
\sum_{ij} G(\mathbf{r},\mathbf{R}_i;k^2)
\left[\hat G+D-\frac{BC}{A+ik}\right]^{-1}_{ij}G(\mathbf{R}_j,\mathbf{R};k^2).
\label{GKrein}
\end{eqnarray}
If $B=C=0$ then the last formula reduces to the so-called ``Krein's resolvent formula'' for the Green function of the system with point-like scatterers \cite{alb88}. Krein's formula can be also obtained by the operator technic of Appendix 1 using the following identities
\begin{eqnarray}
G_p(\mathbf{r},\mathbf{R};k^2)=\langle\mathbf{R}|\frac{1}{H - k^2+W D^{-1} W^+ \mathcal{R}}|\mathbf{r}\rangle=
\nonumber\\\fl=
\langle\mathbf{R}|\frac{1}{H-k^2}-
\frac{1}{H-k^2}W\frac{1}{W^+ \mathcal{R}(H-k^2)^{-1}W + D}W^+\mathcal{R}\frac{1}{H-k^2}|\mathbf{r}\rangle,
\end{eqnarray}
where $|\mathbf{r}\rangle_\mu=\psi_\mu(\mathbf{r}),\,|\mathbf{R}\rangle_\mu=\psi_\mu(\mathbf{R})$.


\end{document}